\documentclass{article}
\usepackage{ijcai20}

\usepackage{times}
\usepackage{soul}
\usepackage{url}
\usepackage[hidelinks]{hyperref}
\usepackage[utf8]{inputenc}
\usepackage[small]{caption}
\usepackage{graphicx}
\usepackage{amsmath}
\usepackage{amsthm}
\usepackage{amsfonts}
\usepackage{booktabs}
\usepackage{algorithm}
\usepackage{algorithmic}
\usepackage{multirow}
\usepackage{caption}
\usepackage{booktabs}
\usepackage{tabularx}
\usepackage{subfigure}

\title{Heterogeneous Graph Collaborative Filtering}

\author{
Zekun Li\thanks{Equal contributions.}$^{1,3}$
\and
Yujia Zheng$^*$$^{2}$\and
Shu Wu$^{4,5}$\and
Xiaoyu Zhang$^{1}$\and
Liang Wang$^{4,5}$
\affiliations
$^1$Institute of Information Engineering, Chinese Academy of Sciences\\
$^2$University of Electronic Science and Technology of China\\
$^3$School of Cyber Security, University of Chinese Academy of Sciences\\
$^4$Center for Research on Intelligent Perception and Computing, Institute of Automation, Chinese Academy of Sciences\\
$^5$School of Artificial Intelligence, University of Chinese Academy of Science
\emails
\{lizekunlee,yjzheng19\}@gmail.com, 
shu.wu@nlpr.ia.ac.cn,
zhangxiaoyu@iie.ac.cn,
wangliang@nlpr.ia.ac.cn}

\begin{document}

\maketitle

\begin{abstract}
Graph-based collaborative filtering (CF) algorithms have gained increasing attention. Existing work in this literature usually models the user-item interactions as a bipartite graph, 
where users and items are two isolated node sets and edges between them indicate their interactions.
Then, the unobserved preference of users can be exploited by modeling high-order connectivity on the bipartite graph. 
In this work, we propose to model user-item interactions as a heterogeneous graph which consists of not only user-item edges indicating their interaction but also user-user edges indicating their similarity.
We develop heterogeneous graph collaborative filtering (HGCF), a GCN-based framework which can explicitly capture both the interaction signal and similarity signal
through embedding propagation on the heterogeneous graph.
Since the heterogeneous graph is more connected than the bipartite graph, the sparsity issue can be alleviated and the demand for expensive high-order connectivity modeling can be lowered.
Extensive experiments conducted on three public benchmarks demonstrate its superiority over the state-of-the-arts.
Further analysis verifies the importance of user-user edges in the graph, justifying the rationality and effectiveness of HGCF.
\end{abstract}

\section{Introduction}
Personalized recommendation plays an important role in many online services such as E-commerce, advertising, and social media. The goal is to estimate how likely a user will adopt an item based on the historical interactions including ratings, clicks, likes, sharing and so on.

Collaborative filtering (CF) is commonly used to leverage these interaction data for recommendation.
It assumes that behaviorally similar users typically share similar interests over items.
Traditional methods focus on measuring the user (item) similarity via various measures such as Jaccard, Cosine and Pearson Correlation~\cite{salton1983introduction,linden2003amazon,hofmann2004latent,resnick1994grouplens,adomavicius2011context}.
Afterwards, a common paradigm for learnable CF models is to transform users and items to low dimensional vectorized representations (aka. embeddings) according to the user-item interactions and thus the learnt embeddings can be used to predict whether a user will interact with an item.
Among those, Matrix Factorization (MF) is a prevalent approach, which models user-item interaction with inner product of their embeddings~\cite{koren2009matrix}.
With the development of deep learning, many methods are proposed to replace the MF interaction function of inner product with nonlinear neural networks~\cite{he2017neural,he2018outer}.
However, these methods only leverage the observed direct interactions between users and items.
The unobserved indirect preference of users latent in the user-item interactions are also crucial for the recommendation.

Recently, graph-based models have gained increasing attention. The user-item interactions are usually modeled as a bipartite graph, where users and items are two isolated node sets and edges between them indicate their interactions. 
Then, the unobserved indirect preference can be captured by modeling high-order connectivity on the bipartite graph.
For example, HOP-Rec performs random walk on the bipartite graph to 
enrich the training data for MF model~\cite{yang2018hop}.
The recently proposed Neural Graph Collaborative Filtering (NGCF) utilizes Graph Convolutional Networks (GCN), an effective framework for representation learning on graphs, to directly inject the high-order graph structure into the learnt embeddings~\cite{wang2019neural}.
Dynamic Graph Collaborative Filtering (DGCF) leverages dynamic graphs to capture sequential relations of items and users for sequential recommendations~\cite{li2020dynamic}.
SMOG-CF, a further proposed GCN-based framework, directly captures the high-order connectivity instead of performing embedding propagation iteratively~\cite{zhangstacked}. 

Despite their effectiveness, we argue that these methods based on the bipartite graph remain two critical limitations.
First, although the crucial user similarity is latent in the bipartite graph structure, these methods lack an explicit modeling of it.
Second, the bipartite graph is sparse with only edges between users and items, which makes it necessary to model an extremely high-order connectivity to capture the indirect preference of users.
However, it is demanding and computationally expensive.
Taking NGCF for example, it relies on GCN to model high-order connectivity by stacking multiple layers. 
Nevertheless, GCN provably performs bad when it is deep with many layers due to repeated Laplacian smoothing~\cite{li2018deeper}.
Therefore, there lacks a guarantee that the extremely high-order connectivity on the bipartite graph, i.e. the indirect preference can be effectively captured.
	
To address these limitations, we construct a heterogeneous graph for recommendation by enriching the bipartite graph with edges between users according to their behavioral similarity.
On the one hand, the crucial user behavioral similarity can be directly encoded in the graph structure and leveraged explicitly in the embedding progress.
On the other hand, the graph becomes more connected and consequently it is easier for a user to reach distant users/items on the graph, which lowers the demand for high-order connectivity modeling.
In addition, the sparsity issue can be alleviated.

\begin{figure}[t]
\centering
\includegraphics[width=0.9\linewidth]{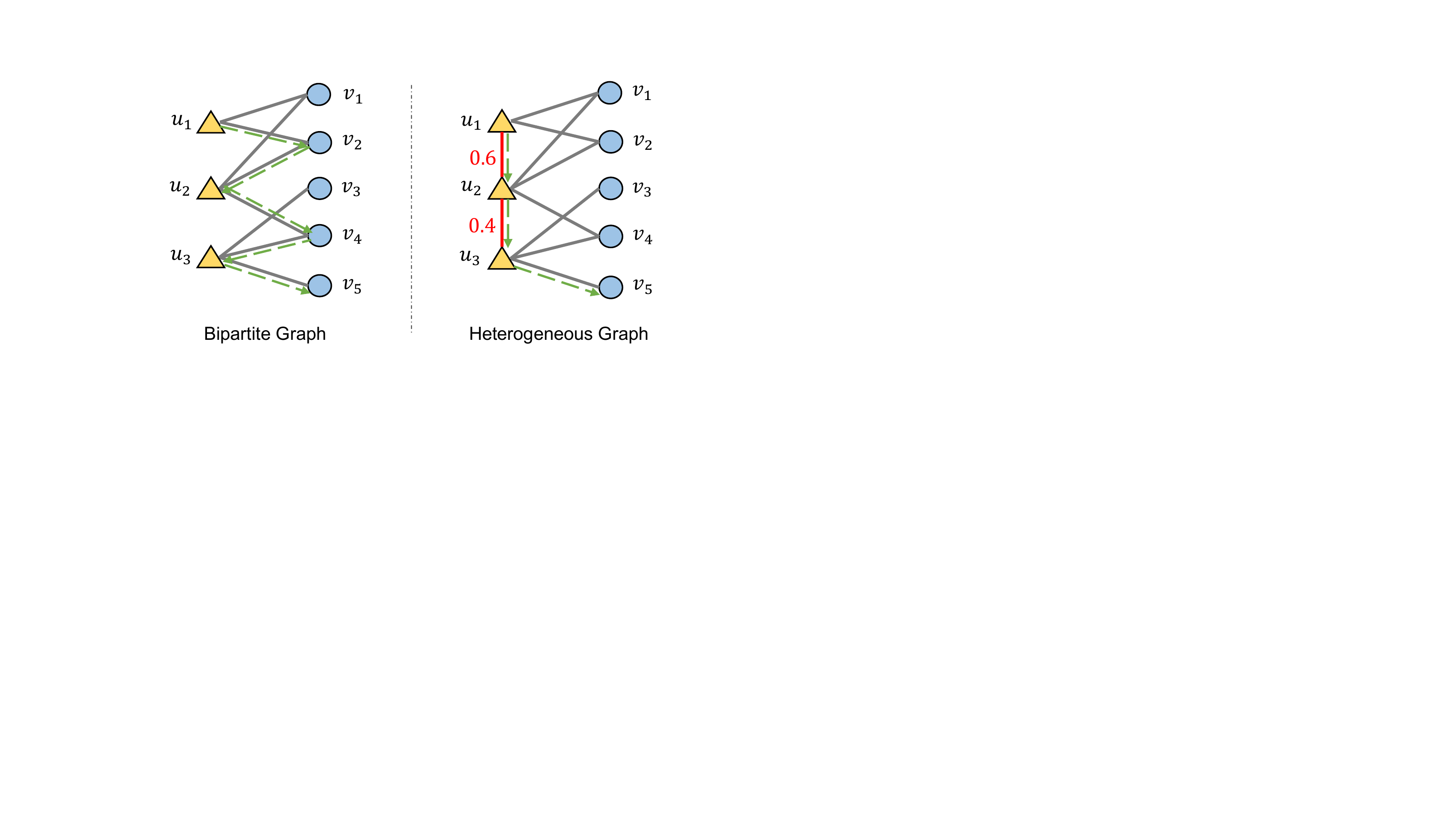}
\caption{An illustration of the bipartite interaction graph (left) and the enriched heterogeneous interaction graph (right). Red lines are enriched edges between users, whose weights are their behavioral similarity.
Green dashed lines are two shortest path from $u_1$ to $u_3$ in the two graphs.}
\label{fig:example}
\vspace{-2mm}
\end{figure}

Figure \ref{fig:example} illustrates a bipartite interaction graph (left) and its enriched heterogeneous interaction graph (right). 
The red lines are enriched edges between users, whose weights are their behavioral similarity. 
Users $u_1$ and $u_2$ are connected since they are behaviorally similar, as they have both interacted with item $v_1$ and $v_2$.
$u_1$ and $u_3$ are not connected since they are not similar enough.
Edge $(u_1, u_2)$ overweights edge $(u_2, u_3)$ since $u_1$ and $u_2$ are more similar than $u_2$ and $u_3$, as they have more neighboring items.  
In this way, the user similarity is explicitly modeled in the graph structure and can be directly leveraged in the embedding progress.
High-order connectivity refers to the path a user reaches any node except its direct neighboring nodes.
For instance, user $u_1$ needs at least 5 steps to reach $v_5$ in the bipartite graph $(u_1 \rightarrow v_2 \rightarrow u_2 \rightarrow v_4 \rightarrow u_3 \rightarrow v_5)$, while in the enriched heterogeneous graph, only 3 steps are enough ($u_1 \rightarrow u_2 \rightarrow u_3 \rightarrow v_4$).
Accordingly, 3-order connectivity modeling is enough to capture the indirect preference of $u_1$ towards $v_5$ in the heterogeneous graph, while 5-order is needed in the bipartite graph. 
Therefore, adding edges between users can make the graph more connected and consequently lower the demand for high-order connectivity modeling.


To better learn embedding of users and items on the heterogeneous graph, we develop a GCN-based framework heterogeneous graph collaborative filtering (HGCF). 
Following the embedding propagation scheme of graph neural networks~\cite{kipf2016semi,hamilton2017inductive,wang2019neural}, we learn the embedding of users/items by aggregating message from their neighboring nodes layer by layer.
The embedding propagation through user-item edges injects the interaction signal into the user/item embeddings, while that through the user-user edges makes the behaviorally similar users closer in the embedding space, whose strength is controlled by the edge weights (similarity). 
Compared with existing methods based on the bipartite graph such as NGCF, our proposed HGCF can capture the indirect user preference on the heterogeneous graph by modeling lower-order connectivity, which is more efficient and can guarantee the embedding performance.
We conduct extensive experiments on three public benchmarks to verify its superiority over the state-of-the-art methods and also the importance of enriched user-user edges in the heterogeneous graph.

Lastly, it is worth mentioning that our proposed HGCF is distinguished from those methods based on Heterogenous Information Network (HIN) or Knowledge Graph (KG), which need extra side information such as social relationships between users and item attributes,
and introduce various types of nodes besides users and items in the graph~\cite{shi2015semantic,wang2017item,shi2018heterogeneous,cao2019unifying,wang2019kgat}.
By contrast, we only need the user-item interaction data, based on which the user similarity can be measured.
The constructed heterogenous graph only contains user, item nodes and user-item, user-user edges.

Overall, our contributions can be summarized as follows:
\begin{itemize}
\item We point out the limitations of existing recommendation methods based on the bipartite interaction graph. 
To address those, we propose to construct a heterogeneous graph which consists of not only user-item edges indicating interaction relations but also edges between users indicating their similarity relation.
\item We develop HGCF, a GCN-based framework which can explicitly capture the user-item interactions and also user similarity through a relative low-order embedding propagation on the enriched heterogeneous graph.
\item Experimental results on three public benchmark datasets demonstrate the superiority of HGCF over the state-of-the-art methods and also the importance of enriched user-user edges in the heterogenous graph. 
\end{itemize}

\begin{figure}[t]
\centering
\includegraphics[width=0.9\columnwidth]{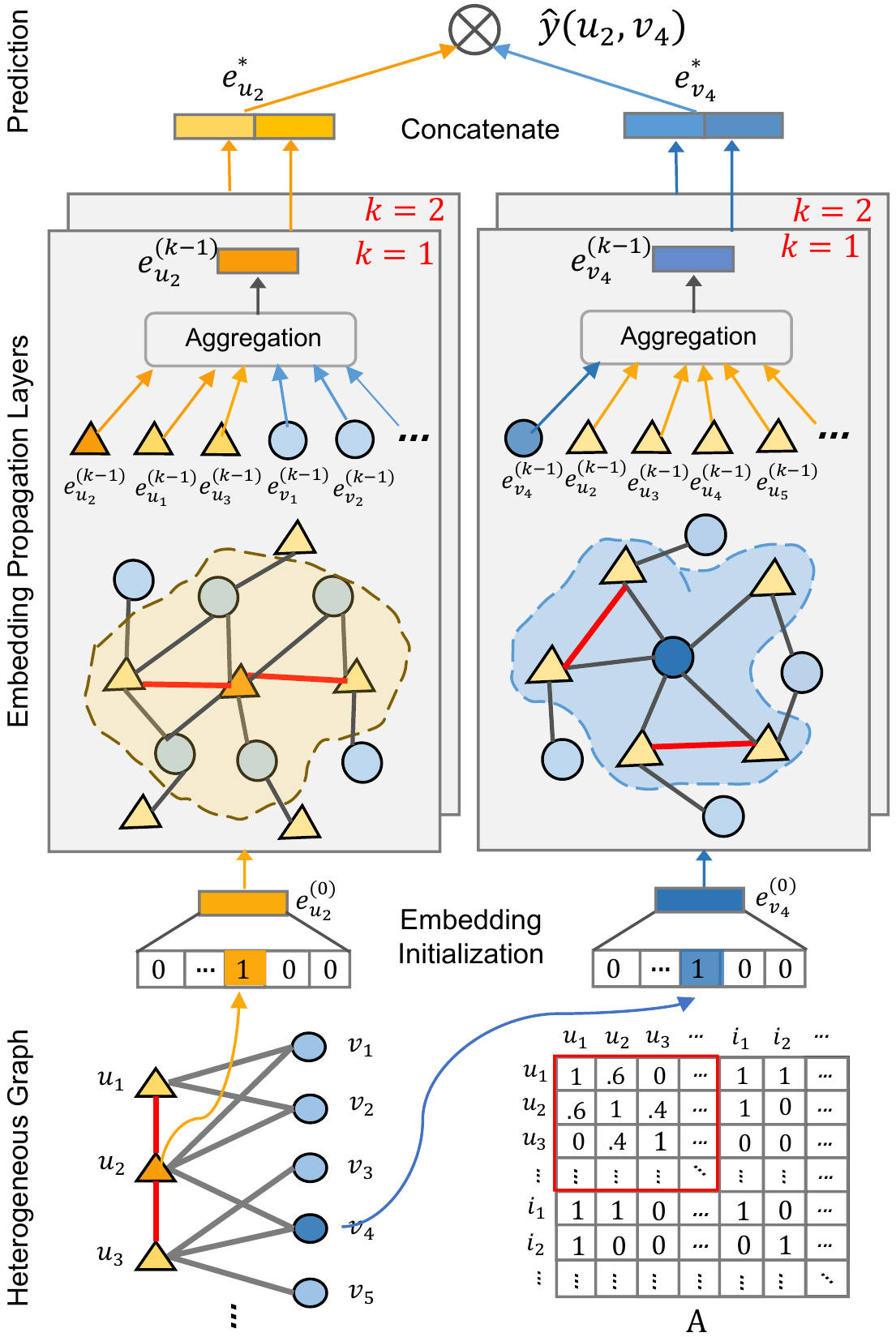} 
\caption{The framework of HGCF. Based on the given user-item interaction data, we construct a heterogeneous graph consisting of user-item edges and user-user edges.  
The embeddings of user $u_2$ and item $v_4$ are first initialized and then refined via multiple embedding propagation layers, whose outputs are concatenated to make the final prediction.}
\label{fig:framework}
\vspace{-4mm}
\end{figure}

\section{Method}
The framework of our proposed method is illustrated in Figure~\ref{fig:framework}.
We first introduce how to model the user-item interaction as a heterogeneous graph and then elaborate the model architecture of HGCF.

\subsection{Heterogeneous Interaction Graph}\label{sect:adj}
In a recommender system, there are $n$ users in $U=\left\{u_1, u_2, \dots, u_n\right\}$ and $m$ items in $V=\left\{v_1, v_2, \dots, v_m\right\}$.
Given the historical interactions between users and items, the goal is to predict how likely a user will interact with an item. 
We build a large heterogeneous graph based on the given user-item interactions.
The heterogeneous graph consists of two types of nodes: users and items, and two types of edges: user-item edges and user-user edges, whose  detailed definition is:

\begin{itemize}
	\item The edge between a user and an item indicates their interaction relation. If a user has interacted with an item, there is an edge between the two nodes, the corresponding position of the adjacency matrix is set to 1.
	\item The edge between two users indicates their behavior similarity relation. Based on the historical user-item interactions, there are many similarity measures can be used, such as Jaccard, Cosine, Euclidean and Pearson Correlation as in the traditional CF methods~\cite{salton1983introduction,linden2003amazon,hofmann2004latent,resnick1994grouplens}.
Nevertheless, in this work, we employ point-wise mutual information (\textbf{PMI}), a popular measure for word associations in Natural Language Processing (NLP) area~\cite{yao2019graph}, to calculate the behavioral similarity between two users. 
We found using PMI achieves better results than using the above mentioned measures in our preliminary experiments. 
Formally, the similarity between a user $u$ and user $u'$, which is also their edge weight, is defined as:
\begin{equation}
\small
\text{PMI}(u, u') = \log \frac{p(u, u')}{p(u)p(u')}
\end{equation}
\begin{equation}
\small
p(u) = \frac{|V(u)|}{|V|}
\end{equation}
\begin{equation}
\small
p(u, u') = \frac{|V(u) \cap V(u')|}{|V|}
\end{equation}
where $V(u)$ denotes the set of items that user $u$ has interacted with, and $V(u) \cap V(u')$ is the set of items that user $u$ and $u'$ have both interacted with. 
We only add the edges with similarity more than a threshold $t$, which is set as 0 by default.
Thus the edges between users that are not similar enough are filtered out.
By this means, we can prevent the matrix from being too dense and the consequent high computational complexity.
\end{itemize}

Besides these two types of edges, we also add the self-loop edges on all the nodes, whose weights are set as 1.
Overall, the whole adjacency matrix of the heterogeneous graph is
\begin{align}
\small
\mathbf{A}_{ij} = 
\begin{cases}
\quad\text{PMI}(i, j)& i,j \text{ are users, } \text{PMI}(i, j) > t\\
\quad 1 & \text{user } i \text{ has interacted with item } j\\
\quad 1 & i = j \\
\quad 0&  \text{otherwise}
\end{cases}
\end{align}
The adjacency matrix $\mathbf{A} \in \mathbb{R}^{(m+n) \times (m+n)}$ consists of both users and items.
An example is illustrated in the lower right corner of Figure~\ref{fig:framework}. The part in red highlighted block contains added user-user edges compared with the bipartite graph.

Analogously, we can also add edges between items which  reflect the item similarity and also densify the graph.
However, our preliminary experiments show that adding item-item edges performs worse than adding user-user edges.
In addition, adding both user-user and item-item edges performs worse than only adding one kind of edges.
Therefore, in this work, besides the user-item edges, we only add user-user edges in our heterogenous graph.
The detailed discuss on this is presented in Section~\ref{sect:graph}.


\subsection{Embedding Initialization} \label{sect:graph}
We describe each user $u$ (item $v$) with an embedding vector $\mathbf{e}_{u}\in\mathbb{R}^{d}$ ($\mathbf{e}_{i}\in\mathbb{R}^{d}$), where $d$ denotes the embedding size.
This can be seen as building a parameter matrix as an embedding look-up table:
\begin{gather}\label{equ:e-0}
\small
	\mathbf{E}=[\underbrace{\mathbf{e}_{u_{1}},\cdots,\mathbf{e}_{u_{n}}}_{\text{users embeddings}}, \underbrace{\mathbf{e}_{v_{1}},\cdots,\mathbf{e}_{v_{m}}}_{\text{item embeddings}}].
\end{gather}

\subsection{Embedding Propagation Layer}
To refine the embeddings of users and items on the heterogeneous interaction graph, we build upon the message-passing architecture of GNNs~\cite{hamilton2017inductive,wang2019neural,hu2020graphair} which has been verified effective in encoding relational information in various applications~\cite{wu2019session,li2019fi,cui2019dressing}.
Each embedding propagation layer takes the former embeddings as input and outputs the updated embeddings which capture higher-order connectivity on the graph.
There are mainly two operations: \emph{message construction} and \emph{message aggregation}.

\noindent{\textbf{Message Construction.~}}
For the $l$-th embedding propagation layer, given a node $i$ (user/item), the message passed from a neighbor $j$ (user/item) is constructed as:
\begin{equation}
\small
\mathbf{m}_{i \leftarrow j}^{(l)} = \widetilde{\textbf{A}}_{ij}(\mathbf{W}_{1}^{(l)}\mathbf{e}_{j}^{(l-1)}+\mathbf{W}_{2}^{(l)}(\mathbf{e}_{i}^{(l-1)} \odot \mathbf{e}_{j}^{(l-1)})).
\end{equation}
$\mathbf{m}_{i \leftarrow j}^{(l)}$ is the message embedding propagated from $j$ to $i$. 
$\mathbf{e}_{i}^{(l-1)}$ and $\mathbf{e}_{j}^{(l-1)}$ are the embeddings of node $i$ and $j$ obtained from the previous layer, where $\mathbf{e}_{i}^{(0)}=\mathbf{e}_{i}$.
$\mathbf{W}_1^{(l-1)},\mathbf{W}_2^{(l-1)}$ are the trainable weight matrices to distill useful information for propagation at the $l$-th layer.
$\widetilde{\textbf{A}}_{ij}$ is the corresponding element in the Laplacian normalized adjacency matrix $\widetilde{\textbf{A}}=\textbf{D}^{-\frac{1}{2}}\textbf{A}\textbf{D}^{-\frac{1}{2}}$, where $\mathbf{D}$ is the diagonal degree matrix.
$\widetilde{\textbf{A}}_{ij}$ controls the propagation strength from $j$ to $i$.
Following NGCF~\cite{wang2019neural}, upon the graph convolutional operation, we add an extra interaction operation between two connected nodes via $\mathbf{e}_{i}^{(l-1)} \odot \mathbf{e}_{j}^{(l-1)}$, where $\odot$ denotes the element-wise product.
This will make the similar nodes pass more messages.

When node $i$ is a user and node $j$ is an item, the message embedding $\mathbf{m}_{i \leftarrow j}^{(l)}$ injects the interaction signal into the embeddings of the user and item, vice versa.
When $i$ and $j$ are both users, $\mathbf{m}_{i \leftarrow j}^{(l)}$ encodes the similarity signal and make them close in the embedding space, whose strength is controlled by their edge weight (similarity) $\widetilde{\textbf{A}}_{ij}$.
When $i=j$, $\mathbf{m}_{i \leftarrow j}^{(l)}$ is the message passed from itself, which helps memorize the message from its ($l$-1)-hop neighbors.
Since it is meaningless to interact with itself, we remove the interaction operation from the message construction on self-loop edges.

It should be noted that we can utilize different trainable weight matrices for different types of edges to better distinguish them.
Nevertheless, we experimentally found that it does not bring about performance improvement, which may attribute to the fact that more parameters make the training harder.

\noindent{\textbf{Message Aggregation.~}}
As we have introduced the constructed message that $i$ receives from the neighbor $j$, we here describe how $i$ aggregates the messages from all its neighbors to refine its embedding.   
Since the users and items have different types of neighboring nodes, we formulate their message aggregation functions respectively for clarity.

For an item $v$, its neighboring nodes are all users, thus it receives messages from neighboring users and itself.
Formally,
\begin{equation}
\small
\mathbf{e}_{v}^{(l)} = \delta(\mathbf{m}_{v \leftarrow v}^{(l)} + \sum_{u \in U(v)}\mathbf{m}_{v \leftarrow u}^{(l)}),
\end{equation}
where $U(v)$ denotes the set of item $v$'s neighboring users and $\delta(\cdot)$ is the LeakyReLU activation function~\cite{maas2013rectifier}.

For a user $u$, there are neighboring users and also neighboring items in the heterogeneous graph, which are denoted as $U(u)$ and $V(u)$, respectively.
It thus receives message from these two types of neighbors and itself:
\begin{equation}
\small
\mathbf{e}_{u}^{(l)} = \delta(\mathbf{m}_{u \leftarrow u}^{(l)} + \sum_{u' \in U(u)}\mathbf{m}_{u \leftarrow u'}^{(l)}+\sum_{v \in V(u)}\mathbf{m}_{u \leftarrow v}^{(l)}).
\end{equation}

It is worth mentioning that we forbid the users to receive message from their neighboring users in the first embedding propagation layer while allowing them to do so in the successive layers.
The intuition is that injecting the interaction signal into the embeddings of users and items is necessary, based on which we can predict the probability that a user interacts with an item.
On the contrary, it is meaningless to make similar users close in the randomly initialized embedding space at the first-order propagation layer.
 

\noindent\textbf{Matrix Form.~}
To offer a holistic view of embedding propagation and facilitate batch implementation, we here provide the matrix form of embedding propagation rule:
\begin{equation}\label{eq:rule}
\small
\mathbf{E}^{(l)}=\delta\Big(\widetilde{\textbf{A}}\mathbf{E}^{(l-1)}\mathbf{W}^{(l)}_{1} + \widetilde{\textbf{A}}\mathbf{E}^{(l-1)}\odot\mathbf{E}^{(l-1)}\mathbf{W}^{(l)}_{2}\Big),
\end{equation}
where $\mathbf{E}^{(l)}$ is the obtained embedding matrix of users and items at the $l$-th layer, where $\mathbf{E}^{(0)}=\mathbf{E}$, that is $\mathbf{e}_{i}^{(0)}=\mathbf{e}_{i}$.
By implementing the matrix-form propagation rule, the embeddings for all users and items in the graph can be simultaneously updated in an efficient manner.

\subsection{Prediction Layer}
After $L$ embedding propagation layers, we can obtain multiple embeddings for a user $u$, namely $\{\mathbf{e}^{(1)}_{u},\cdots,\mathbf{e}^{(L)}_{u}\}$.
Following~\cite{wang2019neural,zhangstacked}, we concatenate them to constitute the final embeddings for users and those for items are obtained in the same way:
\begin{equation}\label{equ:final-rep}
\small
\mathbf{e}^{*}_{u}= \Vert_{l=0,1,\cdots,L}\mathbf{e}^{(l)}_{u}, \quad
\mathbf{e}^{*}_{v}= \Vert_{l=0,1,\cdots,L}\mathbf{e}^{(l)}_{v}, 
\end{equation}
where $\Vert$ is the concatenation operation.
Finally, we conduct the inner product to estimate the preference of a user $u$ towards the target item $v$:
\begin{equation}
\small
\hat{y}(u,v)={\mathbf{e}^{*}_{u}}^\top\mathbf{e}^{*}_{v}.
\end{equation}
A higher $\hat{y}(u,v)$ indicates that user $u$ is more likely to interact with item $v$. 
Various more complicated interaction functions such as multi-layer neural networks can be used, which are left to explore in the future work.

\subsection{Training}
Following~\cite{wang2019neural,zhangstacked}, our adopted loss function consists of two parts:
\begin{gather}\label{equ:loss}
\small
	\mathcal{L} = \mathcal{L}_1+\lambda\mathcal{L}_2,
\end{gather}
where $\mathcal{L}_1$ is the pairwise BPR loss~\cite{rendle2009bpr}, which has been widely used in recommender systems. 
$\mathcal{L}_2$ is the L2 regularization used to avoid overfitting~\cite{zhangstacked}, and $\lambda$ is the trade-off parameter which controls the strength.
Specifically, $\mathcal{L}_1$ and $\mathcal{L}_2$ are defined as:
\begin{equation}\label{equ:loss}
\small
	\mathcal{L}_1 = \sum_{(u,v,v^{-})\in\mathcal{O}}-\ln\sigma(\hat{y}_{uv}-\hat{y}_{uv^{-}}),
\end{equation}
\begin{equation}\label{equ:loss}
\small
	\mathcal{L}_2 = \sum_{(u,v,v^{-})\in\mathcal{O}}\left ( \left \| \mathbf{e}_u \right \| + \left \| \mathbf{e}_v \right \| + \left \| \mathbf{e}_{v^{-}} \right \| \right ),
\end{equation}
where $\mathcal{O}=\{(u,v,v^{-})|(u,v)\in\mathcal{R}^{+}, (u,^{-})\in\mathcal{R}^{-}\}$ denotes the pairwise training data, $\mathcal{R}^{+}$ denotes the positive samples, and $\mathcal{R}^{-}$ is the negative samples.
$\sigma(\cdot)$ is the sigmoid function.
We adopt mini-batch Adam~\cite{kingma2014adam} to optimize the prediction model and update the model parameters.



\section{Experiments}\label{sect:experiments}
In this section, we conduct extensive experiments to answer the following questions:
\begin{itemize}
\item[\textbf{RQ1}]
 How does our proposed HGCF perform compared with the state-of-the-art CF models?
\item[\textbf{RQ2}]
Is the heterogeneous graph better than the bipartite graph for recommendation? What is the best way to construct the graph given the user-item interaction data?
\end{itemize}

\subsection{Experiment Setup}
\noindent\textbf{Datasets.~}
We conduct experiments on the following three datasets:
(1) \textbf{Gowalla} is a dataset containing the users' check-in information on the gowalla, a location-based social network;
(2) \textbf{Yelp2018} is a dataset released at the 2018 Yelp challenge, wherein the locations such as restaurants and theaters are viewed as items; 
(3) \textbf{Amazon-Book} is a subset of the Amazon-review, a widely used dataset for product recommendation~\cite{he2016ups}. 
The statistics of three datasets are summarized in Table~\ref{tab:dataset}.

\begin{table}[t]
\small
\centering\caption{Statistics of evaluation datasets.}
\vspace{-2mm}
\begin{tabular}{|c|c|c|c|} 
\hline
Dataset &\#Users & \#Items & \#Interactions  \\
\hline
Gowalla & 29,858 & 40,981 & 1,027,370 \\
Yelp2018 & 31,668 & 38,048 & 1,561,406 \\
Amazon-Book & 52,643 & 91,599 & 2,984,108 \\
\hline
\end{tabular}\label{tab:dataset}
\vspace{-4mm}
\end{table}

For the three datasets, we only keep users and items with over ten interactions.
We randomly select 80\% in the historical interactions of each user to constitute the training set, and the rest are the test set. 
From the training set we randomly select 10\% as the validation set.
We treat each observed user-item interaction as a positive sample, and randomly select an item to form the negative sample.

\noindent\textbf{Evaluation Metrics.~}
Following~\cite{yang2018hop,wang2019neural}, we adopt two widely-used evaluation metrics: recall@$K$ and ndcg@$K$.
By default, we set $K=20$.
The average metrics for all users in the test set are reported.

\noindent\textbf{Baselines.~}
We compared our proposed HGCF with the following representative methods:
\begin{itemize}
\item \textbf{MF}~\cite{rendle2009bpr} is matrix factorization optimized by the Bayesian personalized ranking (BPR) loss, which only exploits the user-item direct interactions.

\item \textbf{NeuMF}~\cite{he2017neural} first introduces deep neural networks to model the non-linear interactions between users and items.


\item \textbf{CMN}~\cite{ebesu2018collaborative} uses memory networks to enhance user embeddings with those of neighboring users.

\item \textbf{GC-MC}~\cite{berg2017graph} utilizes GCN to generate representations for users and items. Then the recommendation is converted to a link prediction task, where only the first-order neighbors are considered. 

\item \textbf{HOP-Rec}~\cite{yang2018hop} performs random walks to obtain high-order neighbors, which are used to enrich the user-item interaction data.

\item \textbf{PinSage}~\cite{ying2018graph} is originally designed to employ GraphSAGE~\cite{hamilton2017inductive} on item-item graph of pictures on Pinterest. In this work, we apply it on user-item interaction graph. 

\item \textbf{NGCF}~\cite{wang2019neural} utilizes GCN to generate the embedding for users and items via exploring high-order connectivity on the bipartite graph.

\end{itemize}
It should be pointed out that GC-MC, HOP-Rec, PinSAGE, and NGCF are all based on the bipartite graph. 
Among these four methods, GC-MC only leverages the first-order connectivity while the rest three leverage the high-order one.
By contrast, our proposed HGCF explores high-order connectivity on the heterogeneous graph.

\begin{table}
\small
\centering\caption{Performance comparison of different methods on three datasets. Further analysis is provided in Section \ref{sect:result}.
}
\vspace{-2mm}
\setlength{\tabcolsep}{1.5mm}{\begin{tabular}{l|cc|cc|cc} 
\toprule
& \multicolumn{2}{c|}{Gowalla} & \multicolumn{2}{c|}{Yelp2018} & \multicolumn{2}{c}{Amazon-Book}\\
 & recall & ndcg  & recall & ndcg & recall & ndcg\\
\midrule
MF &0.1291 &0.1878 &0.0433 &0.0864 &0.0250 &0.0518 \\  
NeuMF &0.1326 &0.1985 &0.0450 &0.0887 &0.0253 &0.0535\\
CMN &0.1404 &0.2129 &0.0508 &0.0990 &0.0267 &0.0516\\
\hline
GC-MC &0.1395 &0.1960 &0.0462 &0.0923 &0.0288 &0.0551\\
\hline
HOP-Rec &0.1399 &0.2128 &0.0524 &0.0990 &0.0309 &0.0606\\
PinSage &0.1380 &0.1947 &0.0511 &0.0956 &0.0283 &0.0545\\
NGCF &0.1547 &0.2237 &0.0551 &0.1030 &0.0344 &0.0630\\
\hline
HGCF &$\textbf{0.1614}$ &$\textbf{0.2330}$ &$\textbf{0.0572}$ &$\textbf{0.1057}$ &$\textbf{0.0375}$ &$\textbf{0.0671}$\\
\hline \hline
\%Impr. & 4.33\% & 4.15\% & 3.81\% & 2.62\% & 9.01\% & 6.51\% \\ 
\bottomrule
\end{tabular}}
\label{tab:results}
\vspace{-4mm}
\end{table}

\subsection{Model Comparison (RQ1)}\label{sect:result}
The performance of different methods is summarized in Table \ref{tab:results}, from which we can obtain the following observations:
(1) NeuMF consistently outperforms MF on all datasets, proving the importance of non-linearity in modeling interactions between user and item embeddings.
(2) CMN performs better than NeuMF, which suggests that considering the neighboring users can bring about improvement.
In addition, it also outperforms GC-MC, which can be attributed to the attention mechanism.
(3) HOP-Rec achieves great improvement over the two methods CMN and GC-MC, which only leverage the first-order (direct) connectivity. It is reasonable since HOP-Rec explores high-order neighbors to enrich the training data. 
(4) PinSage and NGCF both utilize GCN to inject the high-order connectivity into the embedding progress.
Nevertheless, NGCF outperforms PinSage by a large margin, which may be due to the fact that PinSage only uses the output of the last layer to make prediction while NGCF uses the concatenation of all the layers.
This demonstrates that the embeddings obtained from all the layers have contribution to the prediction.
(5) Our proposed HGCF consistently yields the best performance on all cases.
Specifically, it achieves improvement over the strongest baselines NGCF based on the bipartite graph by 4.33\%, 3.81\% and 9.01\% w.r.t. recall@20 on the three datasets.
This proves the superiority of our constructed heterogeneous graph over the bipartite graph and the rationality of HGCF.


\subsection{Study of HGCF (RQ2)}\label{sect:graph}
To validate and gain deep insights into our proposed method HGCF and the constructed heterogeneous graph, we conduct four groups of experiments.
We first study what is the best way to construct the graph given user-item interaction data, including which types of edges should be added to the graph and which similarity measures should be used for the edge weights.
Then we investigate how the similarity threshold $t$ and layer depth $L$ affect the performance.

\noindent\textbf{Different Edge Types.~}
Given user-item interaction data, we can construct three types of edges as described in Section~\ref{sect:adj}:
(1) \textbf{U-I}: user-item edges indicating interaction relation;
(2) \textbf{U-U}: user-user edges indicating user similarity relation;
(3) \textbf{I-I}: item-item edges indicating item similarity relation, which can be constructed in the same way with user-user edges.
To test which type of edges should be added to the graph besides the necessary user-item edges, we perform experiments over four graphs consisting of different types of edges, and the results are shown in Table~\ref{tab:graph_construction}.

We can observe that adding either user-user edges or item-item edges can improve the performance on Gowalla and Yelp2018 datasets.
On the largest Amazon-Book dataset, the massive item-item edges lead to Out-Of-Memory (OOM) issues. 
The introduction of user-user edges achieves better performance than that of item-item edges on Gowalla and Yelp2018 datasets, which verifies the significance of user similarity.
Surprisingly, adding both user-user edges and item-item edges is inferior to only adding user-user edges.
We conjecture that the simultaneous introduction of two kinds of associations would dilute the interaction signal and lead to embedding space drifting.


\begin{table}[t]
\small
\centering\caption{Performance comparison of different ways to construct the graph. Further analysis is provided in Section \ref{sect:graph}.}
\vspace{-2mm}
\setlength{\tabcolsep}{1.4mm}{\begin{tabular}{l|cc|cc|cc} 
\toprule
& \multicolumn{2}{c|}{Gowalla} & \multicolumn{2}{c|}{Yelp2018} & \multicolumn{2}{c}{Amazon-Book}\\
 & recall & ndcg  & recall & ndcg & recall & ndcg\\
\midrule
UI &0.1547 &0.2237 &0.0551 &0.1030 &0.0344 &0.0630\\
UI+UU &\textbf{0.1614} &\textbf{0.2330} &\textbf{0.0572} &\textbf{0.1057} &\textbf{0.0375} &\textbf{0.0671}\\
UI+II &0.1578 &0.2271 &0.0554 &0.1032 &OOM &OOM\\
UI+UU+II &0.1568 &0.2269 &0.0547 &0.1030 &OOM &OOM\\
\hline
Cosine &0.1554 &0.2240 &0.0557 &0.1035 &0.0365 &0.0652\\
Jaccard &0.1580 &0.2256 &0.0565 &0.1042 &0.0367 &0.0660\\
PMI &$\textbf{0.1614}$ &$\textbf{0.2330}$ &$\textbf{0.0572}$ &$\textbf{0.1057}$ &$\textbf{0.0375}$ &$\textbf{0.0671}$\\
\bottomrule
\end{tabular}}
\label{tab:graph_construction}
\vspace{-4mm}
\end{table}


\noindent\textbf{Different Similarity Measures.~}
We here investigate the influence of different similarity measures on the performance.
Given user-item interaction data, we can measure the weights of edges between users with various similarity measures.
Here we compare our used PMI with two widely used similarity measures in prior work~\cite{adomavicius2011context} \textbf{Cosine} and \textbf{Jaccard}, which can be calculated as: 
\begin{equation}
\small
\text{Cosine}(u, u') = \frac{|V(u)\cap V(u')|}{\sqrt{|V(u)||V(u')|}},
\end{equation}
\begin{equation}
\small
\text{Jaccard}(u, u') = \frac{|V(u)\cap V(u')|}{|V(u)\cup V(u')|}.
\end{equation}
As can be seen in Table~\ref{tab:graph_construction}, using PMI as similarity measures achieves best performance on all cases.
We attribute this to the better distinguishability of PMI since it has a larger scale with both positive and negative values.
On the contrary, Cosine and Jaccard only have positive values in the range of $\left [ 0,1 \right ]$.

\begin{figure}[t]
\centering
\subfigure[Gowalla]{
\begin{minipage}[b]{0.22\textwidth}
\label{fig:hidden}
\includegraphics[width=1\textwidth]{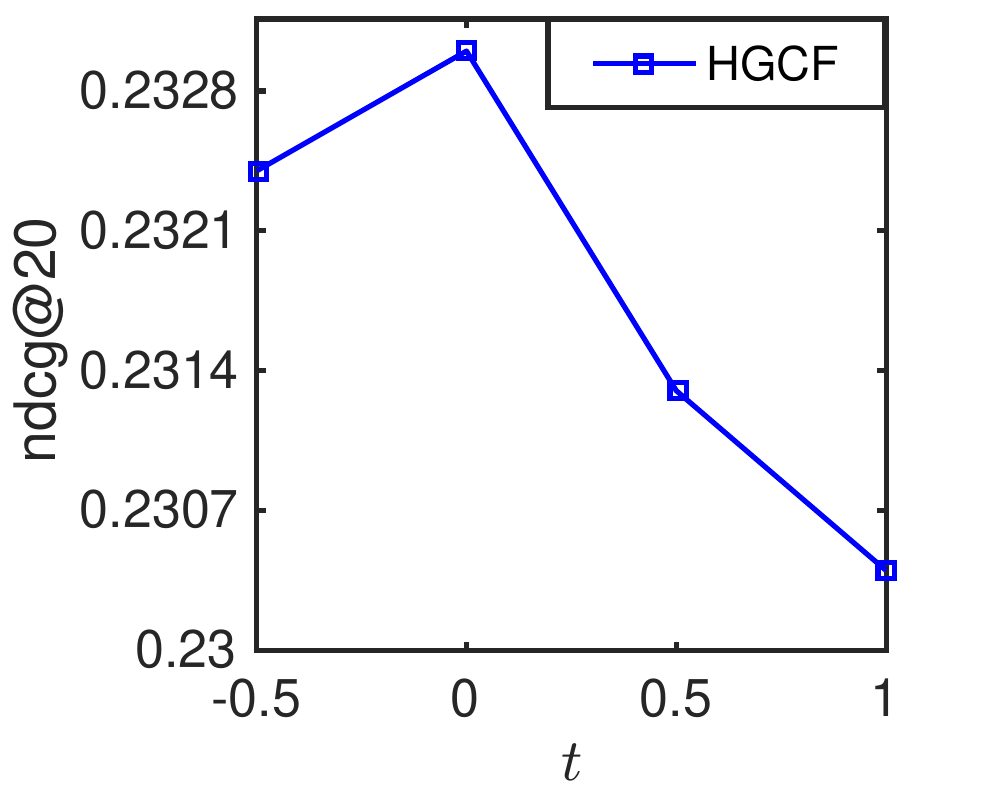}
\includegraphics[width=1\textwidth]{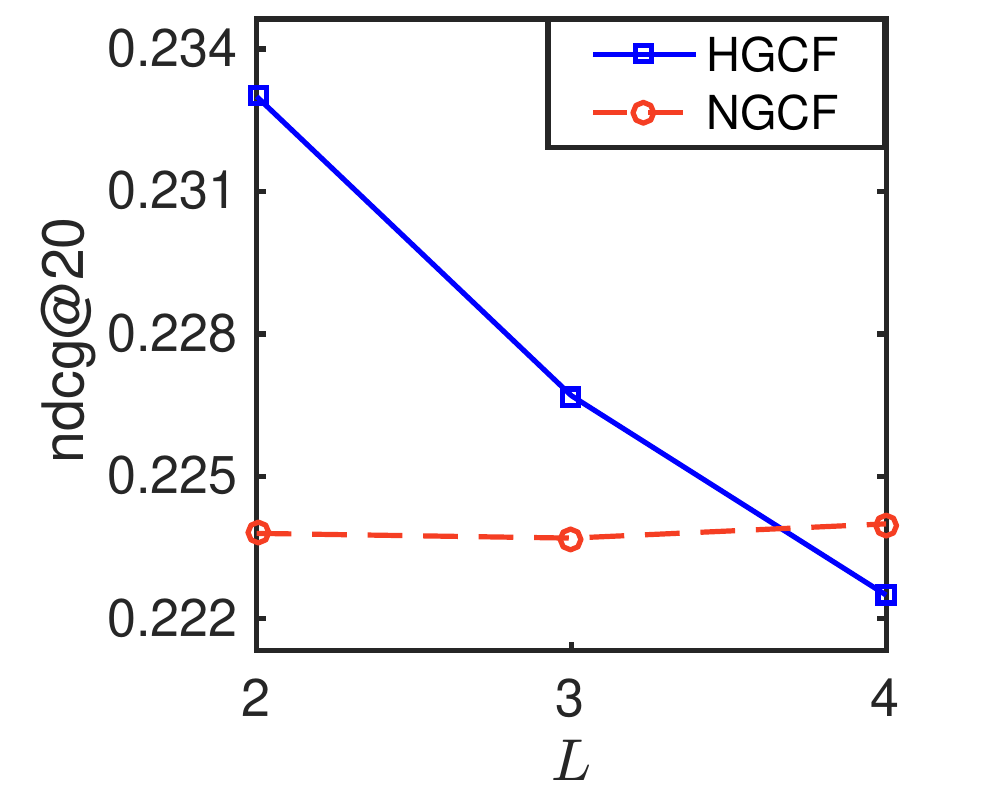}
\end{minipage}%
}%
\subfigure[Yelp2018]{
\begin{minipage}[b]{0.22\textwidth}
\label{fig:order} 
\includegraphics[width=1\textwidth]{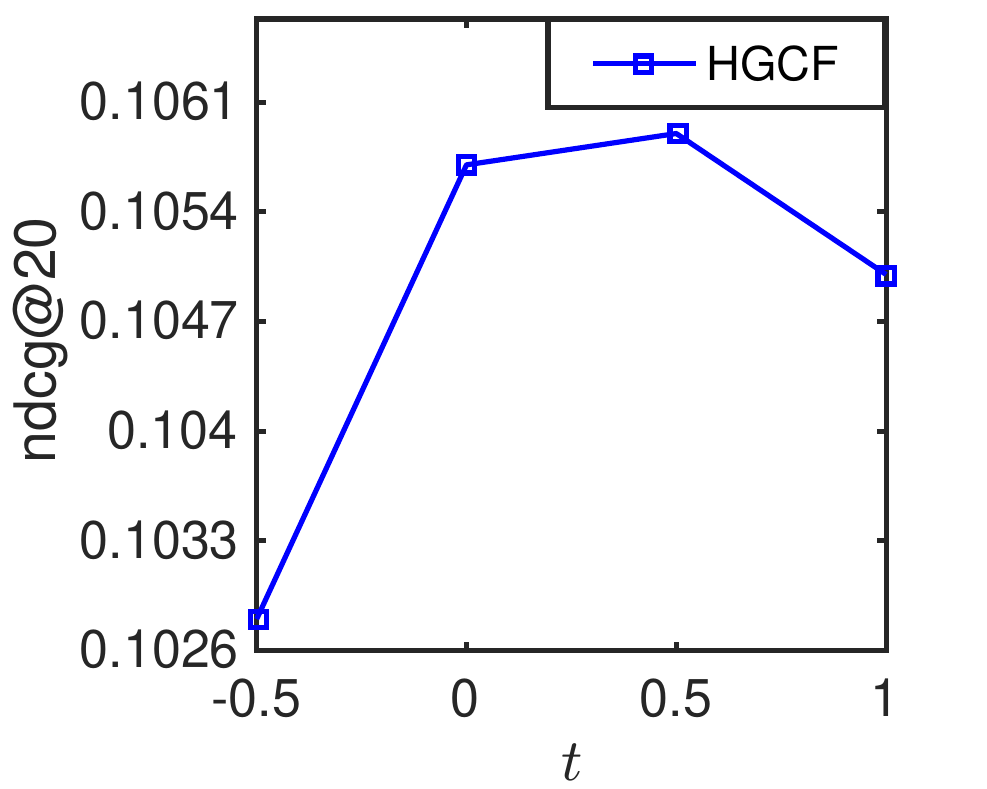}
\includegraphics[width=1\textwidth]{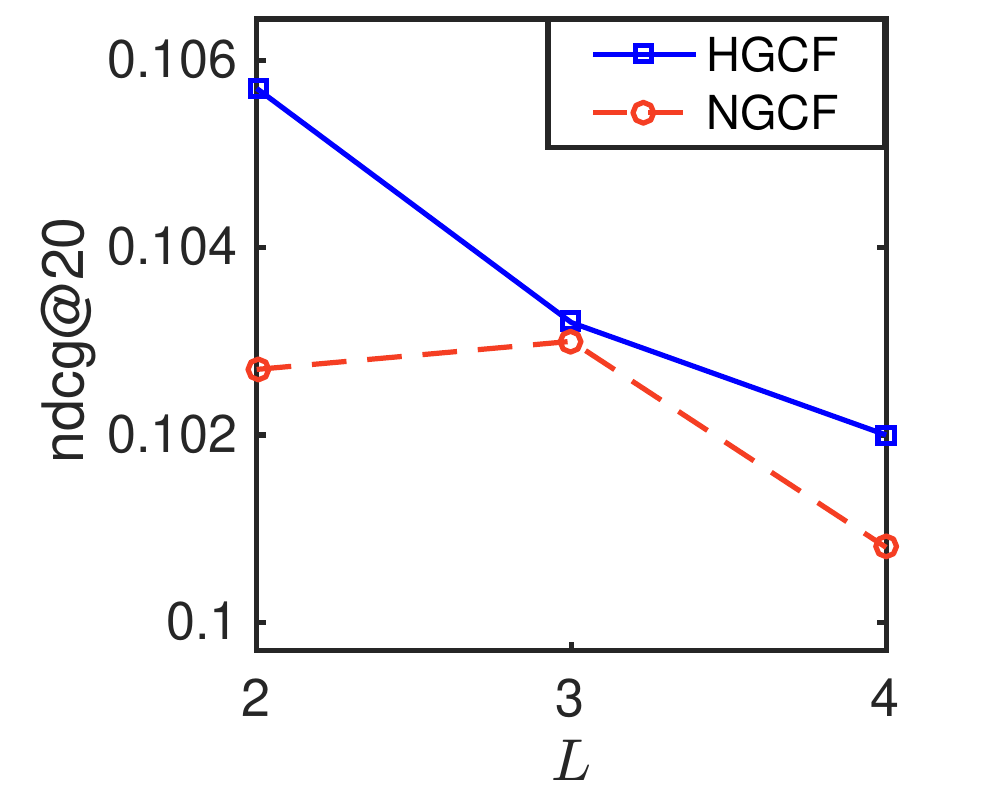}
\end{minipage}%
}%
\vspace{-4mm}
\caption{Performance w.r.t. different similarity threshold $t $ (the upper) and depth of embedding propagation layers $L$ (the lower) on Gowalla and Yelp2018 datasets.}
\label{fig:layer_depth}
\vspace{-5mm}
\end{figure}

\noindent\textbf{Different Similarity Threshold.~}
To investigate the influence of different similarity thresholds on the performance, we vary the threshold $t$ in the range of $\left \{ -0.5, 0, 0.5, 1 \right \}$.
The results are illustrated in the upper of Figure~\ref{fig:layer_depth}.
With the increase of threshold $t$, the graph becomes sparser and the performance first increases and then drops on both datasets.
This indicates that the graph should not be too dense or too sparse, since an extremely dense graph will lead to overfitting issues and an extremely sparse graph fails to fully encode the similarity information.  
Owing to different dataset properties, the performance on Gowalla and Yelp2018 peaks when $t=0$ and $t=0.5$, respectively.
In real applications, we can balance the graph sparsity to obtain the best performance by adjusting the similarity threshold.

\noindent\textbf{Different Layer Depth.~}
As our constructed heterogeneous graph is more connected than the bipartite graph, the indirect preference of users can be captured via lower-order connectivity modeling in theory.
To validate that, we test the performance on bipartite graph (NGCF) and heterogeneous graph (HGCF) varying layer depth, of which the results are shown in the lower of Figure~\ref{fig:layer_depth}.
We can observe that NGCF based on bipartite graph achieves the best performance with three, four layers on Gowalla and Yelp2018 datasets, respectively,
while our proposed HGCF based on heterogeneous graph is the best with only two layers on both datasets.
This proves that the heterogeneous graph can alleviate the demand for high-order connectivity modeling.
However, every coin has two sides.
When there are more than two layers, the performance of HGCF consistently drops.
Nevertheless, it still outperforms NGCF except with four layers on Gowalla.
This is reasonable since the highly-connected graphs are easy to be over-smoothed by multi-layers GCN.
Fortunately, two layers are enough for HGCF to obtain the best performance, which is efficient and can avoid the over-smoothing issues.
Most importantly, HGCF outperforms NGCF by a large margin with less layers, verifying the superiority of our constructed heterogeneous graph and devised HGCF.  

\vspace{-2mm}
 \section{Conclusion}
In this work, we pointed out the limitations of existing CF methods based on bipartite graph.
To address those, we propose to construct a heterogeneous graph which consists of not only user-item edges indicating interaction relations but also edges between users (items) indicating their similarity relation.
We developed HGCF, a GCN-based framework, to explicitly capture the user-item interactions and the user similarity
through embedding propagation on the heterogeneous graph.
Our comparative experiments conducted on three public benchmarks demonstrate its superiority over the state-of-the-art methods.
We also conducted extensive experiments to investigate the best way to construct the graph given user-item interaction data, which may inspire the future work.


\appendix

\bibliographystyle{named}
\bibliography{ijcai20}

\end{document}